\documentclass[comsoc,lettersize,journal]{IEEEtran}

\usepackage{amsmath}
\usepackage{graphicx}
\usepackage{cite}
\usepackage{url}
\usepackage{xcolor}
\usepackage{booktabs}

\usepackage{tikz}
\usetikzlibrary{arrows.meta,positioning,shapes,fit,calc,backgrounds,decorations.pathreplacing, shapes.geometric}

\begin{document}

\title{False-CSI Attacks in Power-Domain NOMA for 6G: A Threat Taxonomy and System-Level Impacts}

\author{Samira~Jafarli, Aysha~Ebrahim, and Suleyman~Uludag%
\thanks{This work has been submitted to the IEEE for possible publication.
Copyright may be transferred without notice, after which this version may no
longer be accessible.}%
\thanks{S. Jafarli is with Young Talents Lyceum of Baku State University, Azerbaijan. }%
\thanks{A. Ebrahim is with the Computer Engineering Department at the University of Bahrain.}%
\thanks{S. Uludag is with the University of Michigan--Flint, Flint, MI, USA
(e-mail: uludag@umich.edu).}}

\markboth{Preprint --- submitted to IEEE Communications Magazine}{Jafarli \MakeLowercase{\textit{et al.}}: False-CSI Attacks in NOMA for 6G}

\maketitle

\begin{abstract}
Power-domain non-orthogonal multiple access (NOMA) remains a widely
studied technique for improving spectral efficiency and supporting dense
connectivity in beyond-5G and 6G networks. Its main operating mechanisms,
however, depend on the integrity of channel-state information (CSI).
Power allocation, user ordering, pairing, clustering, and beamforming can
all be distorted when the CSI consumed by the base station is deliberately
biased rather than merely noisy. This article examines false CSI as an
attack surface in power-domain NOMA. We organize the threat space using a
compact taxonomy with two primary axes: magnitude, which distinguishes
underreporting from overreporting, and ordering effect, which
distinguishes order-preserving, boundary, and order-reversing attacks.
We then show how coordinated false-CSI behavior, group-changing attacks,
direction forgery, pilot spoofing, training-phase injection, and
RIS-induced channel manipulation extend this basic taxonomy. Finally, we
map each attack family to system-level impacts on power allocation, SIC
reliability, scheduler behavior, fairness, throughput, and secrecy. The
central message is that false CSI should be treated not only as a
channel-estimation problem, but also as a control-input integrity problem
for 6G NOMA.
\end{abstract}



\section{Introduction}
\label{sec:intro}

Non-orthogonal multiple access (NOMA) remains an actively studied
multiple-access technique for beyond-5G and 6G networks because it can
improve spectral efficiency and support dense device connectivity in
massive machine-type communications and edge applications. Power-domain NOMA, in
particular, has dominated the security literature because its three
operating mechanisms,  superposition coding, successive interference
cancellation (SIC), and channel-dependent power allocation,  are
familiar enough to analyze rigorously yet coupled enough that
small perturbations propagate. Recent performance studies confirm that
this coupling is also a source of fragility under realistic channel
conditions~\cite{ozduran2024performance}.

The fragility traces back to a CSI-integrity assumption. Every
operational decision a NOMA base station makes,  the decoding order
for SIC, the power split between paired users, the choice of pair or
cluster on a given resource block, and the beam direction in MIMO
variants, is computed from channel information obtained through
reports, feedback, pilot-based estimation, or hybrid acquisition
procedures. The exact CSI path depends on the duplexing mode and system
implementation, but the NOMA scheduler ultimately acts on a channel
input that is assumed to be trustworthy at allocation time. Prior work
has established that the integrity of SIC decoding hinges on the
accuracy of the channel ordering derived from this input
~\cite{ding2020unveiling}. When the input is adversarially biased, the
entire stack can drift with it~\cite{ozduran2024performance}: power
allocations become locally optimal but globally wrong, decoding margins
collapse, and PHY-layer secrecy assumptions may no longer hold.

Most of the literature that touches this dependency treats CSI
imperfection as a stochastic estimation phenomenon, modeled as
zero-mean uncertainty around the true channel. PHY-layer security
surveys have often followed this convention~\cite{pakravan2023physical,
saeed2025comprehensive}, while related attack studies appear separately
under pilot spoofing, untrusted relays, adversarial scheduling, or
malicious reconfigurable intelligent surfaces
(RISs)~\cite{khalid2025malicious,alakoca2022metasurface}. What is still
missing is a compact attack-centered view of the common mechanism behind
these cases: the NOMA scheduler acts on channel information whose
integrity has been biased before power allocation, pairing, decoding
order, or beamforming decisions are made.

This article provides that attack-centered view. Rather than offering
another NOMA performance survey or a countermeasure survey, it treats
false CSI as a threat-modeling problem: scalar misreports, spatial
forgery, training-phase bias, and RIS-induced channel manipulation are
unified by the same NOMA decision chain. Its contributions are three.
First, we organize false-CSI attacks along two axes: a \emph{magnitude}
axis, which distinguishes under-reporting from over-reporting of CSI,
and an \emph{ordering-effect} axis, which distinguishes attacks that
preserve, collapse, or invert the channel ordering used for SIC. Second,
we show how coordinated, group-changing, direction-forgery,
training-phase, and RIS-induced variants extend the basic taxonomy
without changing the underlying dependency on CSI integrity. Third, we
map each attack family to system-level impacts through a propagation
chain that starts at a single corrupted channel input and ends at power
misallocation, SIC failure, scheduler distortion, throughput loss,
fairness degradation, and secrecy leakage. The remainder of the article
is organized as follows: Section~\ref{sec:adversary} frames NOMA's
CSI-driven attack surface; Section~\ref{sec:taxonomy} presents the
taxonomy; Section~\ref{sec:impact} maps the system-level impacts; and
Section~\ref{sec:challenges} discusses implications for 6G NOMA
security.

\section{NOMA in 6G: Why CSI Integrity Matters}
\label{sec:adversary}

\subsection{Power-Domain NOMA in Brief}
\label{sec:noma_basics}

Power-domain NOMA allows two or more users to share the same
time-frequency resource by separating their signals through transmit
power rather than through orthogonal resource blocks. The base station
superposes user signals with different power coefficients. The user with
the weaker channel is typically assigned more power, while the stronger
user receives less power and applies successive interference
cancellation (SIC): it first decodes the weak user's high-power signal,
subtracts it, and then decodes its own lower-power signal from the
residual. The same principle generalizes to clusters of three or more
users, at the cost of additional SIC stages and greater sensitivity to
channel-estimation error~\cite{mohsan2023survey}.

This coupling is attractive for beyond-5G and 6G systems because it can
improve spectral efficiency and support dense connectivity without
requiring new spectrum. It also creates a security-sensitive dependency:
the same channel information that determines who is weak or strong also
determines the power split, SIC order, pairing or clustering decision,
and, in MISO/MIMO-NOMA, the beam direction.

\subsection{Why CSI Is the Soft Underbelly}
\label{sec:csi_underbelly}

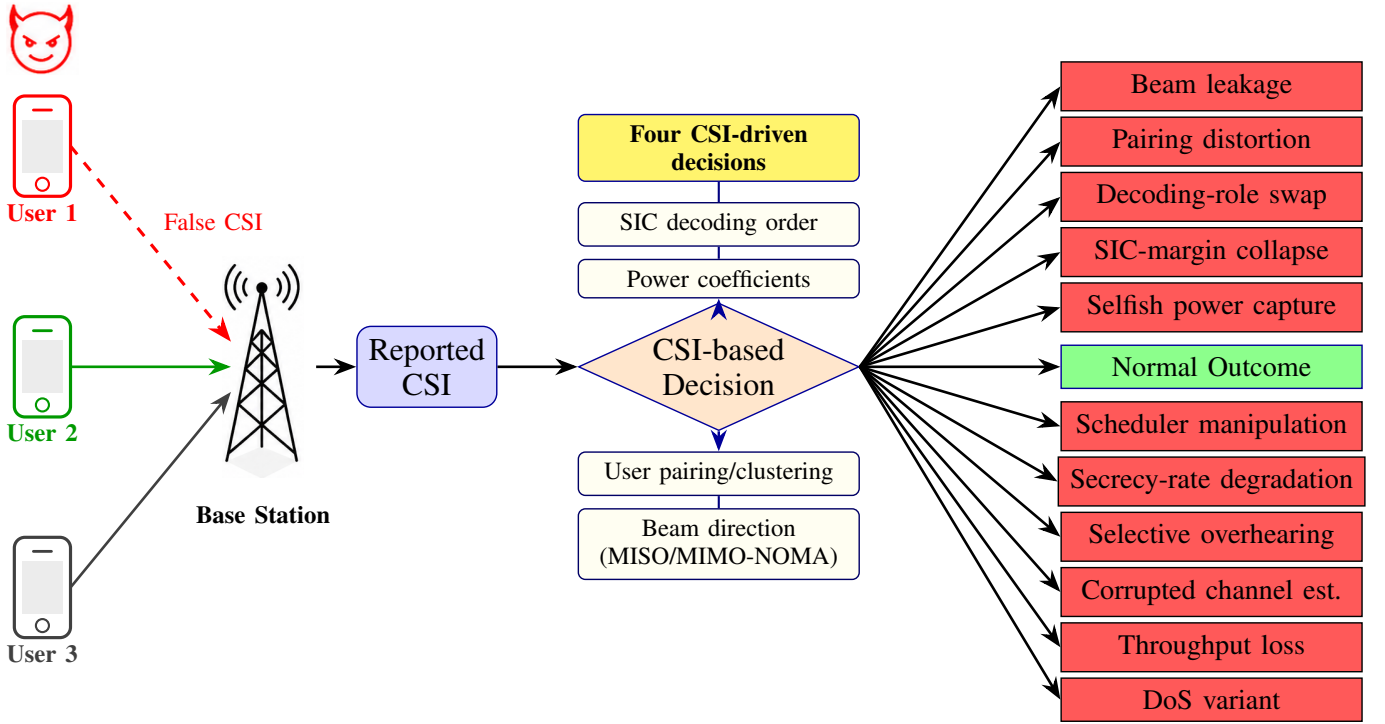
\begin{figure*}[t]
\centering
\resizebox{\textwidth}{!}{%
\begin{tikzpicture}[
    font=\small,
    >=Stealth,
    box/.style={rectangle, draw, rounded corners, align=center,
                minimum width=1.2cm, minimum height=0.75cm},
    decision/.style={diamond, draw, aspect=2.2, align=center,
                     inner sep=1pt, font=\small},
    outbox/.style={rectangle, draw, align=center,
                   minimum width=3.0cm, minimum height=0.35cm,
                   font=\footnotesize},
    arr/.style={->, line width=0.75pt}
]

\tikzset{
  phone/.pic={
    \draw[line width=0.9pt, rounded corners=3pt]
        (-0.28,-0.50) rectangle (0.28,0.50);
    \draw[line width=0.8pt, line cap=round]
        (-0.09, 0.36) -- (0.09, 0.36);
    \draw[line width=0.6pt] (0,-0.38) circle (0.065);
    \fill[gray!15] (-0.20,-0.26) rectangle (0.20,0.26);
  }
}

\begin{scope}[red]
  \pic at (0, 2.2) {phone};
\end{scope}
\node[red,  font=\scriptsize\bfseries] at (0, 1.54) {User 1};
\node[above=2pt] at (0, 2.72)
    {\includegraphics[height=0.72cm]{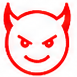}};

\begin{scope}[green!60!black]
  \pic at (0, 0) {phone};
\end{scope}
\node[green!60!black, font=\scriptsize\bfseries] at (0,-0.66) {User 2};

\begin{scope}[black!75]
  \pic at (0,-2.2) {phone};
\end{scope}
\node[black!75, font=\scriptsize\bfseries] at (0,-2.86) {User 3};

\node (bs) at (2.2, 0)
    {\includegraphics[height=2.2cm]{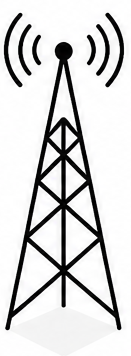}};
\node[font=\scriptsize\bfseries, below=1pt of bs] {Base Station};


\draw[arr, red, dashed, line width=0.9pt]
    (0.28, 2.2) -- (1.88, 0.25)
    node[pos=0.50, above right, font=\scriptsize, red] {False CSI};
 
\draw[arr, green!60!black]
    (0.28, 0.0) -- (1.88, 0.0);
 
\draw[arr, black!75]
    (0.28,-2.2) -- (1.88,-0.25);

\node[box,      right=0.4cm of bs, fill=blue!15, draw=blue!60!black] (csi) {Reported\\CSI};
\node[decision, right=0.8cm of csi, fill=orange!20, draw=blue!60!black] (dec) {CSI-based\\Decision};

\node[above=1.20cm of dec,
      draw=blue!60!black, fill=yellow!70, rounded corners=2pt,
      font=\scriptsize\bfseries, align=center,
      minimum width=2.8cm, inner sep=3pt]
    (hdr) {Four CSI-driven\\decisions};

\node[below=0.20cm of hdr,
      draw=blue!40!black, fill=yellow!6, rounded corners=2pt,
      font=\scriptsize, align=center,
      minimum width=2.8cm, inner sep=3pt]
    (d1) {SIC decoding order};

\node[below=0.12cm of d1,
      draw=blue!40!black, fill=yellow!6, rounded corners=2pt,
      font=\scriptsize, align=center,
      minimum width=2.8cm, inner sep=3pt]
    (d2) {Power coefficients};

\node[below=0.20cm of dec,
      draw=blue!40!black, fill=yellow!6, rounded corners=2pt,
      font=\scriptsize, align=center,
      minimum width=2.8cm, inner sep=3pt]
    (d3) {User pairing/clustering};

\node[below=0.12cm of d3,
      draw=blue!40!black, fill=yellow!6, rounded corners=2pt,
      font=\scriptsize, align=center,
      minimum width=2.8cm, inner sep=3pt]
    (d4) {Beam direction\\{\scriptsize(MISO/MIMO-NOMA)}};

\draw[arr] (bs.east) -- (csi.west);
\draw[arr] (csi) -- (dec);

\draw[arr, blue!60!black, line width=0.7pt]
    (dec.north) -- (d2.south);
\draw[-, blue!50!black, line width=0.6pt]
    (d2.north) -- (d1.south);
\draw[-, blue!50!black, line width=0.6pt]
    (d1.north) -- (hdr.south);

\draw[arr, blue!60!black, line width=0.7pt]
    (dec.south) -- (d3.north);
\draw[-, blue!50!black, line width=0.6pt]
    (d3.south) -- (d4.north);

\node[outbox, right=2.0cm of dec, fill=green!45, draw=blue!60!black]       (normal) {Normal Outcome};
\node[outbox, above=0.12cm of normal, fill=red!65]   (a1) {Selfish power capture};
\node[outbox, above=0.05cm of a1, fill=red!65]       (a2) {SIC-margin collapse};
\node[outbox, above=0.05cm of a2, fill=red!65]       (a3) {Decoding-role swap};
\node[outbox, above=0.05cm of a3, fill=red!65]       (b2) {Pairing distortion};
\node[outbox, above=0.05cm of b2, fill=red!65]       (c1) {Beam leakage};
\node[outbox, below=0.12cm of normal, fill=red!65]   (b1) {Scheduler manipulation};
\node[outbox, below=0.05cm of b1, fill=red!65]       (c2) {Secrecy-rate degradation};
\node[outbox, below=0.05cm of c2, fill=red!65]       (c3) {Selective overhearing};
\node[outbox, below=0.05cm of c3, fill=red!65]       (d1) {Corrupted channel est.};
\node[outbox, below=0.05cm of d1, fill=red!65]       (d2) {Throughput loss};
\node[outbox, below=0.05cm of d2, fill=red!65]       (d3) {DoS variant};

\foreach \x in {normal,a1,a2,a3,b1,b2,c1,c2,c3,d1,d2,d3}
    \draw[arr] (dec.east) -- (\x.west);

\end{tikzpicture}%
}
\caption{False-CSI attack surface in power-domain NOMA. User~1 sends
biased channel information, while Users~2 and~3 are legitimate. The base
station's CSI-driven decisions can produce normal operation or one of
several adversarial effects, which are organized in
Section~\ref{sec:taxonomy} and mapped to system-level impacts in
Section~\ref{sec:impact}.}
\label{fig:noma_stack}
\end{figure*}

Every NOMA allocation begins with CSI. As shown in
Fig.~\ref{fig:noma_stack}, the base station uses channel information to
choose the SIC decoding order, power coefficients, user pairing or
clustering, and beam direction in MISO/MIMO-NOMA. This makes CSI more
than a performance variable: it is a control input.

This article uses \emph{false CSI} as an umbrella term for
adversarially biased channel information consumed by the NOMA decision
logic. The biased input may be a reported channel gain, a channel-quality
indicator, a spatial channel direction, or an upstream channel estimate
corrupted during training. We do not assume that the base station is
physically blind; rather, we assume that the scheduling and allocation
logic acts on CSI whose integrity may be biased before the NOMA decision
is made.

This is the structural difference from orthogonal multiple access. In a
TDMA or OFDMA system, a false channel report may mainly affect the
reporting user's own allocation. In NOMA, the same false input also
affects the paired user's decoding role, the shared power split, and the
cluster or beam used by nearby users. The next subsection captures this
dependency through the SIC margin that false-CSI attacks manipulate.

\subsection{The SIC Decoding Margin}
\label{sec:sic_condition}

For a two-user NOMA pair, let the near user have the stronger channel
and the far user have the weaker channel. The far user is typically
assigned more transmit power, while the near user applies SIC by
decoding and subtracting the far user's signal before decoding its own.
The exact SIC feasibility condition depends on the receiver, coding and
modulation choices, channel model, and residual-interference assumptions.
For the taxonomy below, it is sufficient to use a simplified received
power-gap condition as an illustrative SIC margin~\cite{ding2020unveiling}:
\begin{equation}
\bigl(P_{\mathrm{far}} - P_{\mathrm{near}}\bigr)\,|h_{\mathrm{near}}|^2
\;\geq\; \zeta,
\label{eq:sic_condition}
\end{equation}
where $P_{\mathrm{far}}$ and $P_{\mathrm{near}}$ are the allocated
powers, $|h_{\mathrm{near}}|^2$ is the near user's channel power gain,
and $\zeta$ denotes an effective SIC-margin threshold.

False CSI does not violate \eqref{eq:sic_condition} directly. Instead,
it causes the base station to choose the wrong near--far roles or the
wrong power split before the inequality is evaluated. Three attack
effects follow: the false CSI may preserve the ordering but skew the
power allocation, push two users to a boundary where SIC becomes
fragile, or reverse the ordering so that the SIC-margin check is satisfied
for the wrong assignment. These three effects define the
ordering-effect axis of the taxonomy below.

\section{Threat Taxonomy of False-CSI Attacks}
\label{sec:taxonomy}

\subsection{The Two Axes}
\label{sec:two_axes}

A useful taxonomy of false-CSI attacks must be compact enough for a
magazine article but precise enough to distinguish attack mechanics,
affected NOMA decisions, and system-level consequences. We use two axes,
summarized in Fig.~\ref{fig:taxonomy}. The first is \emph{magnitude}:
the adversary can understate or overstate the relevant channel value.
The second is \emph{ordering effect}: the false CSI can preserve the
reported channel ordering, push users to a boundary where ordering
becomes ambiguous, or reverse the ordering used for SIC. Crossing these
axes yields six base cases. Coordinated, group-changing, spatial, and
training-phase attacks then extend these cases without changing the
central dependency: NOMA decisions are computed from corrupted CSI.

\begin{figure*}[t]
\centering
\begin{tikzpicture}[
    font=\sffamily\small,
    >=Latex,
    line width=0.8pt,
    cell/.style={
        draw=black!50, rounded corners=2pt,
        minimum width=4.0cm, minimum height=1.55cm,
        align=center, inner sep=4pt, text width=3.7cm
    },
    cellU/.style={cell, fill=green!8, draw=green!50!black},
    cellO/.style={cell, fill=blue!8,  draw=blue!50!black},
    rowlabel/.style={
        rounded corners=2pt,
        minimum width=2.6cm, minimum height=1.55cm,
        align=center, font=\bfseries\small, text width=2.4cm
    },
    rowlabelU/.style={rowlabel, fill=green!18, draw=green!50!black},
    rowlabelO/.style={rowlabel, fill=blue!18,  draw=blue!50!black},
    rowlabelMod/.style={rowlabel, fill=violet!16, draw=violet!60!black,
        minimum height=1.45cm},
    rowlabelOrth/.style={rowlabel, fill=orange!16, draw=orange!70!black,
        minimum height=1.45cm},
    collabel/.style={
        align=center, font=\bfseries\small,
        minimum width=4.0cm, text width=3.7cm
    },
    modifier/.style={
        draw=violet!60!black, fill=violet!10, rounded corners=2pt,
        align=center, font=\small, minimum width=4.0cm,
        minimum height=1.45cm, text width=3.7cm, inner sep=4pt
    },
    orthogonal/.style={
        draw=orange!70!black, fill=orange!8, rounded corners=2pt,
        align=center, font=\small, minimum width=4.0cm,
        minimum height=1.45cm, text width=3.7cm, inner sep=4pt
    }
]

\node[
    draw=black!55, fill=gray!14, rounded corners=2pt,
    minimum width=2.6cm, minimum height=0.62cm,
    align=center, font=\bfseries\small
] at (-3.5, 3.10) {Attack type};
\node[
    draw=black!55, fill=gray!14, rounded corners=2pt,
    minimum width=12.8cm, minimum height=0.62cm,
    align=center, font=\bfseries\small
] at (4.2, 3.10) {Ordering effect $\longrightarrow$};

\node[collabel] at ( 0.0, 2.20) {Order-preserving\\\scriptsize(stealthy)};
\node[collabel] at ( 4.2, 2.20) {Boundary\\\scriptsize(SIC-margin collapse)};
\node[collabel] at ( 8.4, 2.20) {Order-reversing\\\scriptsize(label swap)};

\node[rowlabelU] (rU) at (-3.5, 0.70)
    {Under-\\reporting\\\scriptsize\textnormal{(strong $\to$ weaker)}};
\node[cellU] at (0.0, 0.70)
    {Selfish power capture\\\scriptsize\S\,III.B $\cdot$ Fig.\,3 (1A)};
\node[cellU] at (4.2, 0.70)
    {SIC margin collapses;\\high outage probability\\\scriptsize\S\,III.B $\cdot$ Fig.\,3 (1B)};
\node[cellU] at (8.4, 0.70)
    {Decoding-role swap;\\weak user forced to SIC\\\scriptsize\S\,III.B $\cdot$ Fig.\,3 (1C)};

\node[rowlabelO] (rO) at (-3.5,-0.95)
    {Over-\\reporting\\\scriptsize\textnormal{(weak $\to$ stronger)}};
\node[cellO] at (0.0,-0.95)
    {Scheduler / ranking\\manipulation\\\scriptsize\S\,III.C $\cdot$ Fig.\,4 (2A)};
\node[cellO] at (4.2,-0.95)
    {SIC margin collapses;\\tied ordering\\\scriptsize\S\,III.C $\cdot$ Fig.\,4 (2B)};
\node[cellO] at (8.4,-0.95)
    {Decoding-role swap;\\attacker fails own SIC\\\scriptsize\S\,III.C $\cdot$ Fig.\,4 (2C)};

\draw[decorate, decoration={brace, amplitude=6pt}, line width=0.9pt, black!60]
    ($(rU.north west)+(-0.16, 0.0)$) -- ($(rO.south west)+(-0.16, 0.0)$);
\node[rotate=90, font=\bfseries\small, anchor=south]
    at ($(rU.west)!0.5!(rO.west)+(-0.76,0)$) {Magnitude};

\node[rowlabelMod] (rM) at (-3.5,-2.65)
    {Cross-\\cutting\\\scriptsize\textnormal{(overlays)}};
\node[modifier] at (0.0,-2.65)
    {\textbf{Group-changing}\\\scriptsize Crosses clustering boundary;\\forces re-pairing};
\node[modifier] at (4.2,-2.65)
    {\textbf{Coordinated}\\\scriptsize Colluding users mask\\individual bias};
\node[modifier] at (8.4,-2.65)
    {\textbf{Applies to any cell}\\\scriptsize Overlays, not separate families};

\node[rowlabelOrth] (rT) at (-3.5,-4.25)
    {Orthogonal\\\scriptsize\textnormal{(points of\\injection)}};
\node[orthogonal] at (0.0,-4.25)
    {\textbf{Direction forgery}\\\scriptsize Beam leakage; RIS-induced bias (\S\,III.D)};
\node[orthogonal] at (4.2,-4.25)
    {\textbf{Training-phase injection}\\\scriptsize Pilot spoofing / false CSI (\S\,III.D)};
\node[orthogonal] at (8.4,-4.25)
    {\textbf{Layer onto any cell}\\\scriptsize Combine with magnitude attacks};

\end{tikzpicture}
\caption{Threat taxonomy of false-CSI attacks in power-domain NOMA. The
\emph{magnitude} axis captures whether the relevant CSI value is
understated or overstated; the \emph{ordering-effect} axis captures
whether the reported ordering is preserved, collapsed to a boundary, or
reversed. The six base cells describe scalar CSI attacks, while
group-changing and coordinated behavior act as overlays. Direction
forgery and training-phase injection are separate points of CSI
corruption that can combine with the base cases.}
\label{fig:taxonomy}
\end{figure*}

The two axes are intentionally operational. Magnitude is the lever the
adversary controls; ordering effect is the NOMA decision consequence.
In the order-preserving case, the base station keeps the same near--far
roles but computes the wrong power split. In the boundary case, the
reported gains become too close for a stable SIC margin. In the
order-reversing case, the base station assigns near--far roles to the
wrong users and therefore satisfies the SIC condition for the wrong
power assignment.

The remaining rows of Fig.~\ref{fig:taxonomy} clarify scope. A
\emph{group-changing} attack crosses a pairing or clustering threshold
and changes which users share a resource block. A \emph{coordinated}
attack distributes bias across multiple users, making the aggregate
allocation error larger than any one false report. Direction forgery and
training-phase injection sit outside the scalar magnitude axis, but they
corrupt the same control input: the CSI used by the base station before
NOMA allocation, beamforming, or SIC decisions are made.

\subsection{Underreporting: Strong-to-Weaker False CSI}
\label{sec:underreporting}

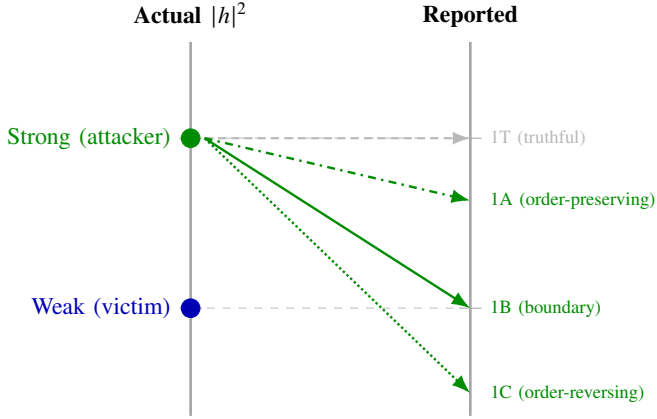
\begin{figure}[t]
\centering
\begin{tikzpicture}[
    x=3.7cm, y=0.75cm,
    font=\sffamily\small,
    >=Latex,
    arr/.style={->, line width=0.9pt, green!55!black}
]
\draw[line width=1.0pt, gray!70] (0,-1.4) -- (0,5.2);
\draw[line width=1.0pt, gray!70] (1,-1.4) -- (1,5.2);
\foreach \y in {0.5, 3.5} {
    \draw[gray!70] (-0.04,\y) -- (0.04,\y);
    \draw[gray!70] ( 0.96,\y) -- (1.04,\y);
}
\node[font=\bfseries\small, above=2pt] at (0,5.2) {Actual $|h|^2$};
\node[font=\bfseries\small, above=2pt] at (1,5.2) {Reported};
\draw[gray!40, dashed, line width=0.5pt] (0,3.5) -- (1,3.5);
\draw[gray!40, dashed, line width=0.5pt] (0,0.5) -- (1,0.5);

\filldraw[green!55!black] (0,3.5) circle (3.5pt);
\node[left=4pt, green!55!black, font=\small, anchor=east] at (0,3.5)
    {Strong (attacker)};
\filldraw[blue!70!black] (0,0.5) circle (3.5pt);
\node[left=4pt, blue!70!black, font=\small, anchor=east] at (0,0.5)
    {Weak (victim)};

\draw[->, line width=0.8pt, gray!55, densely dashed]
    (0.05,3.5) -- (1,3.5);
\node[right=4pt, gray!55, font=\scriptsize] at (1,3.5) {1T (truthful)};

\draw[arr, dash dot]      (0.05,3.5) -- (1,2.4);
\node[right=4pt, green!55!black, font=\scriptsize] at (1,2.4)
    {1A (order-preserving)};

\draw[arr]                (0.05,3.5) -- (1,0.5);
\node[right=4pt, green!55!black, font=\scriptsize] at (1,0.5)
    {1B (boundary)};

\draw[arr, densely dotted](0.05,3.5) -- (1,-1.0);
\node[right=4pt, green!55!black, font=\scriptsize] at (1,-1.0)
    {1C (order-reversing)};
\end{tikzpicture}
\caption{Underreporting by a strong user. Lowering the reported channel
gain can preserve the ordering, collapse the users to a boundary, or
reverse the near--far roles used for SIC.}
\label{fig:strong_misreport}
\end{figure}

A central scalar false-CSI attack is a strong-channel user claiming a
weaker channel than it has. NOMA assigns more power to weaker users, so
an attacker that successfully appears weaker captures a larger share of
the transmit power. Three subcases follow from how aggressively the
attacker shrinks its reported gain (Fig.~\ref{fig:strong_misreport}).

In the \emph{order-preserving} case
(Fig.~\ref{fig:strong_misreport}, case~1A), the attacker reports a value
that is lower than its true gain but still above its victim's. The base
station's ordering of users is unchanged, and so is the SIC decoding
sequence; only the size of the power gap is distorted. This is the
stealthiest variant because nothing in the BS's view of the system looks
abnormal: the attacker remains the strong user, the victim remains the
weak user, and the SIC decoding margin of
Section~\ref{sec:sic_condition} continues to be satisfied. Only the rate
split has shifted in the attacker's favor.

In the \emph{boundary} case
(Fig.~\ref{fig:strong_misreport}, case~1B), the attacker reports a gain
at or near its victim's level. The two users are now indistinguishable to
the ordering rule, and the power separation is insufficient to support
reliable SIC. The decoding margin collapses, residual interference
accumulates, and outage probability rises sharply for both users in the
pair.

In the \emph{order-reversing} case
(Fig.~\ref{fig:strong_misreport}, case~1C), the attacker reports a gain
below its victim's. The BS now treats the attacker as the weak user and
grants it the larger power coefficient. Because the attacker's true
channel is strong, it decodes its own signal easily despite being
assigned the role of weak user, while the genuine weak user is forced
into a decoding position its channel cannot support.

These cases also admit group-changing variants. If the false report
crosses a pairing or clustering threshold, the scheduler may move the
attacker into a different NOMA pair or cluster, creating resource-block
churn beyond the original two-user interaction. Coordinated versions are
discussed in Section~\ref{sec:coordinated}.

\subsection{Overreporting: Weak-to-Stronger False CSI}
\label{sec:overreporting}

\begin{figure}[t]
\centering
\begin{tikzpicture}[
    x=3.7cm, y=0.75cm,
    font=\sffamily\small,
    >=Latex,
    arr/.style={->, line width=0.9pt, blue!70!black}
]
\draw[line width=1.0pt, gray!70] (0,-0.6) -- (0,5.6);
\draw[line width=1.0pt, gray!70] (1,-0.6) -- (1,5.6);
\foreach \y in {0.5, 3.5} {
    \draw[gray!70] (-0.04,\y) -- (0.04,\y);
    \draw[gray!70] ( 0.96,\y) -- (1.04,\y);
}
\node[font=\bfseries\small, above=2pt] at (0,5.6) {Actual $|h|^2$};
\node[font=\bfseries\small, above=2pt] at (1,5.6) {Reported};
\draw[gray!40, dashed, line width=0.5pt] (0,3.5) -- (1,3.5);
\draw[gray!40, dashed, line width=0.5pt] (0,0.5) -- (1,0.5);

\filldraw[green!55!black] (0,3.5) circle (3.5pt);
\node[left=4pt, green!55!black, font=\small, anchor=east] at (0,3.5)
    {Strong (victim)};
\filldraw[blue!70!black] (0,0.5) circle (3.5pt);
\node[left=4pt, blue!70!black, font=\small, anchor=east] at (0,0.5)
    {Weak (attacker)};

\draw[->, line width=0.8pt, gray!55, densely dashed]
    (0.05,0.5) -- (1,0.5);
\node[right=4pt, gray!55, font=\scriptsize] at (1,0.5) {2T (truthful)};

\draw[arr, dash dot]      (0.05,0.5) -- (1,1.8);
\node[right=4pt, blue!70!black, font=\scriptsize] at (1,1.8)
    {2A (order-preserving)};

\draw[arr]                (0.05,0.5) -- (1,3.5);
\node[right=4pt, blue!70!black, font=\scriptsize] at (1,3.5)
    {2B (boundary)};

\draw[arr, densely dotted](0.05,0.5) -- (1,5.0);
\node[right=4pt, blue!70!black, font=\scriptsize] at (1,5.0)
    {2C (order-reversing)};
\end{tikzpicture}
\caption{Overreporting by a weak user. Raising the reported channel gain
can manipulate ranking decisions, collapse the SIC boundary, or reverse
the near--far roles assigned by the base station.}
\label{fig:weak_misreport}
\end{figure}
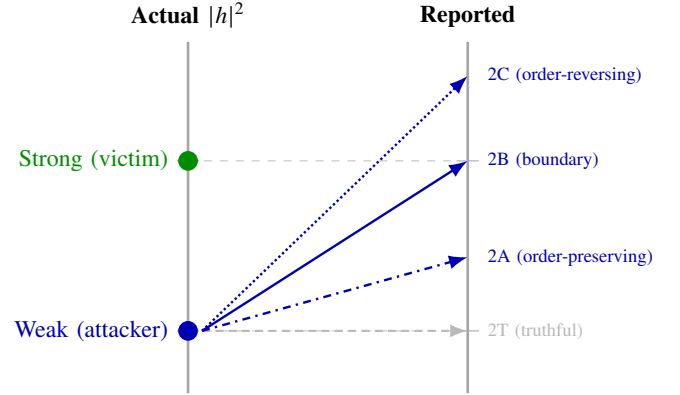

Overreporting is the symmetric case: a user claims a better channel than
it actually has. The mechanics mirror Section~\ref{sec:underreporting},
but the incentives differ. A weak user that overreports gets a smaller
power share under NOMA, which directly hurts its own rate. The attack
therefore rarely makes sense as pure rate-grabbing; instead, overreporting
is most effective when the goal is to manipulate \emph{ranking-driven}
decisions such as user selection, proportional-fair scheduling priority,
beam pointing, or pairing assignment. The same three ordering effects
recur (Fig.~\ref{fig:weak_misreport}).

In the \emph{order-preserving} case
(Fig.~\ref{fig:weak_misreport}, case~2A), the weak user's report is
inflated but still below the strong user's true gain. The pair ordering
is preserved, but a scheduler that combines NOMA with proportional
fairness or quality-of-service ranking may treat the attacker as more
deserving of resources than its actual channel justifies, capturing
service that would have gone to other users in the cell.

In the \emph{boundary} case
(Fig.~\ref{fig:weak_misreport}, case~2B), the inflated report ties the
strong user. As with boundary underreporting, the SIC decoding margin
collapses, but the responsibility is now reversed: it is the attacker's
report that pulls the gap closed.

In the \emph{order-reversing} case
(Fig.~\ref{fig:weak_misreport}, case~2C), the weak user reports a gain
above the strong user's. The BS now treats the weak user as the strong
one and assigns it the smaller power coefficient, while the genuine
strong user is assigned the larger share and is expected to decode
without applying SIC. The configuration is doubly wrong: the attacker,
with a weak true channel, cannot reliably perform the SIC subtraction it
has been assigned, and the genuine strong user receives no decoding
instructions suited to its actual channel.

A strong user can also overreport, exaggerating an already favorable
channel; this rarely changes scalar ordering but may still bias
ranking-driven scheduling. Spatial manipulation of beam direction is
treated separately in Section~\ref{sec:beyond_magnitude}.

\subsection{Spatial and Training-Phase False CSI}
\label{sec:beyond_magnitude}

The six cells in Fig.~\ref{fig:taxonomy} describe scalar false CSI: the
attacker biases a reported or estimated channel value upward or
downward. 6G NOMA also depends on non-scalar and pre-report channel
information. We include those attacks here not as new taxonomy cells, but
as additional \emph{points of CSI corruption} that feed the same NOMA
decision pipeline.

\textbf{Spatial false CSI} corrupts the direction of the channel rather
than its magnitude. In MISO and MIMO-NOMA, the base station uses channel
direction to form beams and spatial nulls. A malicious user may report a
direction correlated with a victim's channel so that the resulting beam
leaks energy toward the attacker. A compromised or malicious RIS can
produce a similar effect by changing the apparent spatial channel
observed at the base station, even without user-side
collusion~\cite{khalid2025malicious,alakoca2022metasurface}.

\textbf{Training-phase false CSI} corrupts the channel estimate before
the NOMA scheduler acts. Pilot contamination, pilot spoofing, or
training-phase injection can bias the estimate used for ordering, power
allocation, pairing, clustering, or beamforming~\cite{nashat2025statistical}.
Reactive training jamming is the denial-of-service version: instead of
biasing the estimate toward a favorable allocation, it prevents the base
station from obtaining a usable estimate.

Thus, spatial and training-phase attacks do not expand the taxonomy
horizontally. They specify \emph{where} false CSI enters the system,
while the magnitude and ordering-effect axes specify \emph{how} the
corrupted input changes NOMA decisions.

\subsection{Coordinated and Group-Changing Attacks}
\label{sec:coordinated}

Every attack family above admits a coordinated extension in which two or
more malicious users combine their reports. Coordination changes the
attack surface in three ways. First, it makes the boundary and
order-reversing cases of
Sections~\ref{sec:underreporting} and~\ref{sec:overreporting} far easier
to engineer: a single user reporting near a victim's gain may be flagged
as a statistical outlier, but two users reporting consistent near-victim
values look like a plausible pair. Second, coordination opens attacks
that no single user can mount alone, such as $N$-attacker spread
underreporting in which a small bias is distributed across many users so
that no individual report appears anomalous, while the aggregate effect
on power allocation is large. Third, in direction forgery, coordinated
multi-attacker schemes can reconstruct enough of the victim's spatial
signature to enable selective eavesdropping that single-attacker schemes
cannot.

The same logic applies to the training phase: cooperative pilot spoofing
distributes the false-pilot burden across several attackers, defeating
detectors that rely on energy or spatial signatures of a single source.

Two features of 6G amplify the coordinated threat. Ultra-dense
deployments raise the prior probability that any given pair of users
includes at least one compromised endpoint~\cite{son2024adversarial},
and history-aware scheduling widens the window over which a coordinated
bias can act. Coordinated false-CSI behavior therefore deserves explicit
treatment in any 6G threat model, not relegation to a footnote on the
single-attacker case.

\section{System-Level Impact}
\label{sec:impact}

\begin{figure*}[t]
\centering
\begin{tikzpicture}[
    font=\sffamily\small,
    >=Latex,
    line width=0.8pt,
    stage/.style={
        draw=black!60, rounded corners=3pt,
        minimum width=2.7cm, minimum height=1.45cm,
        align=center, font=\small\bfseries,
        text width=2.5cm, inner sep=4pt
    },
    s1/.style={stage, fill=red!12,    draw=red!60!black},      
    s2/.style={stage, fill=orange!14, draw=orange!70!black},   
    s3/.style={stage, fill=yellow!18, draw=yellow!50!black},   
    s4/.style={stage, fill=blue!10,   draw=blue!60!black},     
    s5/.style={stage, fill=violet!12, draw=violet!60!black},   
    consequence/.style={
        font=\scriptsize\itshape,
        align=center, text width=2.5cm,
        gray!50!black
    },
    bigarrow/.style={
        ->, line width=1.4pt, gray!60!black,
        shorten >=2pt, shorten <=2pt
    },
    feedback/.style={
        ->, line width=0.9pt, red!50!black, dashed,
        shorten >=2pt, shorten <=2pt
    }
]

\node[s1] (cause)  at ( 0.0, 0) {False CSI\\\scriptsize(\S\,III)};
\node[s2] (alloc)  at ( 3.6, 0) {Power\\allocation\\distortion};
\node[s3] (sic)    at ( 7.2, 0) {SIC margin\\violation};
\node[s4] (link)   at (10.8, 0) {Throughput\\\&\ fairness\\loss};
\node[s5] (secy)   at (14.4, 0) {Secrecy\\leakage\\(if MIMO/RIS)};

\draw[bigarrow] (cause)  -- (alloc);
\draw[bigarrow] (alloc)  -- (sic);
\draw[bigarrow] (sic)    -- (link);
\draw[bigarrow] (link)   -- (secy);

\node[consequence, above=0.25cm of alloc] {wrong inputs $\Rightarrow$\\locally optimal,\\globally wrong};
\node[consequence, above=0.25cm of sic]
   {Eq.~\eqref{eq:sic_condition}\\satisfied for the\\wrong assignment};
\node[consequence, above=0.25cm of link]
   {block-error rate\\and HARQ\\storms};
\node[consequence, above=0.25cm of secy]
   {selective\\overhearing of\\targeted users};

\node[consequence, below=0.25cm of alloc] {scheduler\\symptom:\\low};
\node[consequence, below=0.25cm of sic]   {PHY-layer\\symptom:\\medium};
\node[consequence, below=0.25cm of link]  {link-layer\\symptom:\\high};
\node[consequence, below=0.25cm of secy]  {silent\\secrecy\\leakage};

\draw[feedback] (link.south) ++(0,-1.65)
    -- ++(-7.2,0)
    -| (cause.south);
\node[font=\scriptsize\itshape, red!55!black]
   at (7.2,-2.3) {hybrid-ARQ feedback corrupts next-slot ranking};

\end{tikzpicture}
\caption{Impact-propagation chain for a false-CSI attack. A single
corrupted channel input cascades through power allocation, SIC decoding,
link-layer throughput and fairness, and, in MIMO/RIS-assisted NOMA,
secrecy. The lower labels indicate where the attack first appears as an
operational symptom: scheduler bias may be subtle, SIC failure is more
visible, and secrecy leakage can remain silent. Hybrid-ARQ feedback
closes a loop back to next-slot scheduling, making the perturbation
persistent.}
\label{fig:impact_chain}
\end{figure*}
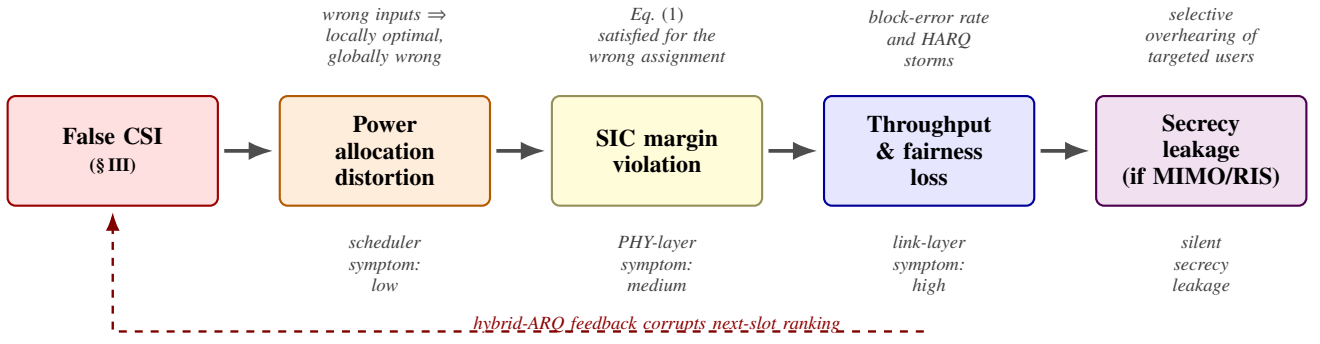

The taxonomy of Section~\ref{sec:taxonomy} describes what an attacker
\emph{does}; this section describes what \emph{breaks}. The impacts
propagate through a coupled chain (Fig.~\ref{fig:impact_chain}), and
their severity depends jointly on the magnitude axis of the attack and
on which downstream stage absorbs the corrupted input.
Table~\ref{tab:attack_comparison} maps the four attack families onto
their manipulated CSI input, primary system-level effect, and the point
at which the attack becomes operationally visible.

\begin{table*}[t]
\centering
\caption{False-CSI attack families in power-domain NOMA, organized by
manipulated CSI input, primary system effect, layer of impact, and
attack visibility.}
\label{tab:attack_comparison}
\renewcommand{\arraystretch}{1.22}
\setlength{\tabcolsep}{5.5pt}
\small
\begin{tabular}{@{}p{3.0cm} p{3.1cm} p{4.25cm} p{2.75cm} p{1.85cm}@{}}
\toprule
\textbf{Attack family} &
\textbf{Manipulated CSI} &
\textbf{Primary system effect} &
\textbf{Layer of impact} &
\textbf{Attack visibility} \\
\midrule
Underreporting (\S\,III.B)
& Reported channel gain $|h|^2$
& Selfish power capture; SIC-margin collapse; decoding-role swap
& Scheduler, PHY, link
& Low (1A) to high (1C) \\
\addlinespace[2pt]
Overreporting (\S\,III.C)
& Reported channel gain $|h|^2$
& Scheduler/ranking manipulation; pairing distortion; possible role swap
& Scheduler, PHY
& Medium \\
\addlinespace[2pt]
Spatial false CSI (\S\,III.D)
& CSI direction; RIS-induced apparent channel
& Beam leakage; secrecy-rate degradation; selective overhearing
& Beamforming, air interface
& Low / silent \\
\addlinespace[2pt]
Training-phase false CSI (\S\,III.D)
& Pilot signal or sounding-phase estimate
& Corrupted channel estimate; throughput loss; denial-of-service variant
& PHY, scheduler
& Medium to high \\
\bottomrule
\end{tabular}
\end{table*}

\textbf{Power allocation distortion and pairing errors.}
The base station's power-allocation routine is a constrained optimizer
that takes CSI values as input and returns quasi-optimal allocations.
When the inputs are biased, the optimizer still produces a locally
optimal solution \emph{with respect to the CSI it received}, which is
what makes the distortion difficult to distinguish from a legitimately
unfavorable channel realization. The misallocation manifests as both an
efficiency loss, aggregate spectral efficiency falls below the
truthful baseline,  and an unfairness, since power redirected to
attackers is taken from genuinely weak users whose channel information
is accurate. The same corrupted input feeds the pairing and clustering
stage: false CSI near a clustering threshold can move an attacker into a
different NOMA pair or cluster, leading to scheduler churn that ripples
across the cell. Coordinated false-CSI behavior amplifies this effect by
making the perturbation persistent rather than
transient~\cite{son2024adversarial}, and dense 6G deployments compound
it further because each biased CSI input can affect more potential pair
partners than in 5G.

\textbf{SIC failure and error propagation.}
Once power allocation is misaligned with the true channels, the SIC
decoding margin of \eqref{eq:sic_condition} is checked against the wrong
quantities. In the boundary case from
Section~\ref{sec:underreporting}, the BS satisfies the inequality on
paper while the genuine post-channel power gap is at or below the
sensitivity floor. The strong user's first-stage decoding then fails to
cleanly subtract the weak user's signal, residual interference leaks
into the second decoding stage, and outage probability rises sharply
for both users.At the link layer this surfaces as elevated block-error rates,
retransmission storms, and adaptive-modulation fallback to lower-rate
codes; effects that can resemble ordinary channel-quality
degradation rather than an adversarially biased CSI input. The order-reversing case is more
severe: the user the BS expects to perform SIC is the one with the weak
true channel and cannot reliably decode the high-power signal it has
been told to subtract. Errors propagate not just within the pair but
to the ranking inputs used by the scheduler in the next slot, since
hybrid-ARQ feedback and channel-quality reports both depend on the
broken decoding chain.

\textbf{Throughput, fairness, and quality of service.}
The aggregate effect on cell throughput depends on how persistently
false CSI biases scheduling and power allocation. Order-preserving
variants are especially damaging at scale because they have low
operational visibility: a small steady bias across many users can shift
a fairness-weighted rate distribution toward attackers while still
resembling natural channel variation. Selfish-user equilibria may emerge
in which truthful reporting becomes individually disadvantageous, and
quality-of-service guarantees for genuine weak users degrade because
resources nominally reserved for them have been redirected. Over time,
this hollowing-out of QoS becomes a form of system instability: the
scheduler may report that service targets are being met while genuine
weak users experience persistent under-service. The 6G targets of
$10^6$ devices/$\mathrm{km}^2$ and ultra-reliable low-latency
communications make this regime particularly fragile because fairness
must hold across very large user populations on tight timescales.

\textbf{Confidentiality and secrecy leakage.}
The direction-forgery and beam-leakage subfamilies of
Section~\ref{sec:beyond_magnitude} act on a different layer: rather than
distorting the rate split, they bend the spatial pattern of the
transmission so that signals intended for legitimate users illuminate
the
attacker~\cite{khalid2025malicious,alakoca2022metasurface}. The result
is a degradation of secrecy rate that classical eavesdropping models do
not predict, because the eavesdropper here is also an authenticated user
whose feedback the system relies on. Coordinated direction forgery
sharpens the leakage further by allowing several attackers to triangulate
a victim's spatial signature, enabling selective overhearing of the
specific users targeted by the scheduler.

\section{Implications for 6G NOMA Security}
\label{sec:challenges}

The attack families above point to a broader lesson for 6G NOMA:
false CSI is not merely a channel-estimation nuisance. It is a
control-input integrity problem. Because the same CSI values determine
ordering, power allocation, pairing, clustering, beamforming, and SIC
behavior, even a small adversarial bias can propagate across multiple
layers of the system. Five implications follow.

\textbf{False CSI should be modeled as adversarial input.}
Most NOMA studies treat imperfect CSI as random estimation uncertainty.
That model is useful for performance analysis, but it does not capture
strategic behavior by selfish or malicious users. In a false-CSI attack,
the distortion is not zero-mean noise; it is chosen to alter a specific
decision made by the base station. A strong user may underreport to
capture power, a weak user may overreport to manipulate scheduling, and
a coordinated group may distribute small biases across many reports so
that no single report appears suspicious. Attack-aware NOMA analysis
therefore needs an explicit adversarial-CSI profile, not only a
statistical error model.

\textbf{Dense deployment amplifies small reporting biases.}
The 6G vision of ultra-dense connectivity changes the scale of the
problem. A biased CSI input in a sparse system affects a limited number of
pairings; in a dense NOMA cell, the same false-CSI value can alter which
users are clustered, how resource blocks are shared, and which users
inherit difficult SIC responsibilities. The danger is therefore
not only the size of one false report, but the number of downstream
decisions that depend on it. At the scale of massive machine-type
communications, many low-amplitude false reports may be more damaging
than a single obvious outlier.

\textbf{Hybrid attacks blur the boundary between PHY-layer and control
logic.}
The taxonomy shows that false CSI can enter through several points:
reported channel gain, spatial channel direction, uplink training, pilot
spoofing, or RIS-induced manipulation. These vectors are often studied
separately, but in a 6G NOMA system they affect the same decision chain.
A pilot-spoofing attack can bias the channel estimate before allocation;
a direction-forgery attack can bend the beam pattern after user
selection; and a malicious or compromised RIS can change the apparent
channel seen by the base station~\cite{khalid2025malicious,
alakoca2022metasurface}. Treating these as isolated attack classes risks
missing the combined effect of corrupted CSI on scheduling, power
allocation, SIC, and secrecy.

\textbf{AI-driven scheduling increases the value of manipulation.}
Future 6G systems are expected to rely increasingly on learning-based or
history-aware resource allocation. This creates an additional incentive
for stealthy false-CSI behavior. An attacker does not need to cause an
immediate outage to gain an advantage; it can bias the scheduler's view
of channel quality, reliability, or service demand over time.
Adversarial behavior at higher 6G layers can therefore compound
false-CSI attacks at the NOMA layer~\cite{son2024adversarial}. The most
dangerous cases may be those that remain below short-term anomaly
thresholds while gradually changing long-term scheduling decisions.

\textbf{Standardized adversarial-CSI evaluation is missing.}
The final implication is methodological. There is no common benchmark
for comparing false-CSI attacks in NOMA: different papers use different
channel models, attacker objectives, user densities, pairing rules, and
CSI-error assumptions. This makes it difficult to compare attack severity
across underreporting, overreporting, direction forgery, pilot spoofing,
and RIS-induced manipulation. A useful benchmark would specify the
attacker's knowledge, the manipulated CSI field, the intended NOMA
decision error, and the observable system-level consequences. Such a benchmark would also connect naturally to emerging 6G trust and
zero-trust frameworks. Recent surveys of trust evaluation and 6G
security~\cite{saeedi2024trust,scalise2024systematic} and reference
zero-trust architectures~\cite{itrust6g,trinh2025framework,wang2023six}
offer a foundation, but the benchmark's first purpose should be clear
attack modeling rather than defense comparison.

\section{Conclusion}
\label{sec:conclusion}

False CSI creates a distinctive attack surface in power-domain NOMA
because the same channel input shapes user ordering, power allocation,
pairing and clustering, beamforming, SIC reliability, fairness,
throughput, and secrecy. This article treated false CSI as an
adversarial control-input problem rather than as ordinary channel
estimation error. We organized the resulting attacks along two axes:
\emph{magnitude}, which distinguishes underreporting from
overreporting, and \emph{ordering effect}, which distinguishes attacks
that preserve, collapse, or reverse the channel ordering used for SIC.

The central lesson is that a false channel input may appear local, but
its consequences are system-wide. A small bias can alter the power
split, weaken the SIC margin, distort scheduler behavior, shift
resources away from truthful users, or create silent secrecy leakage in
MIMO/RIS-assisted settings. For 6G-scale NOMA, CSI integrity should
therefore be treated as part of the attack model itself: before
resource allocation, trust management, or evaluation methodology can be
made robust, the field needs a clear understanding of how false CSI
enters the system and how its effects propagate.

\bibliographystyle{IEEEtran}
\bibliography{nomaRefs}

\vspace{-1.35cm}
\begin{IEEEbiographynophoto}{Samira~Jafarli}
Samira Jafarli studies at the Young Talents Lyceum of Baku State University, Azerbaijan. Her research interests include next-generation wireless communications, 6G networks, artificial intelligence, cybersecurity, and intelligent infrastructure systems. She received a Gold Medal at the GENIUS Olympiad 2025 and a Second Place award in the National Selection Tour for the International Science and Engineering Fair (ISEF) for her engineering and scientific research projects. Her current work focuses on the security and reliability of future communication networks and the application of emerging technologies to real-world challenges.  
\end{IEEEbiographynophoto}

\vspace{-1.35cm}

\begin{IEEEbiographynophoto}{Aysha~Ebrahim}
(Senior Member, IEEE) received her B.Sc. degree in Computer Engineering with first class honors in 2009 from the University of Bahrain. She completed her M.Sc. degree, with distinction, in Electronic Engineering from the University of York in 2011. In 2016, she received her Ph.D. degree in Electrical and Electronic engineering from the University of Manchester. Aysha is currently working in the University of Bahrain as an assistant professor in the department of computer engineering. 
Aysha is a Fellow of the UK Higher Education Academy (HEA), a member of the Institute of Electrical and Electronic Engineers (IEEE), a board member in IEEE ComSoc Bahrain chapter.  In 2019, she received the best paper award in the prestigious IEEE Wireless Communication and Networking Conference (IEEE WCNC 2019).
Her research focuses on 5G and beyond wireless networks, MAC layer design for wireless communication systems, green wireless networking, interference and radio resource management.
\end{IEEEbiographynophoto}

\vspace{-1.35cm}

\begin{IEEEbiographynophoto}{Suleyman~Uludag}
Suleyman Uludag is the David M. French Professor of Computer Science in the
College of Innovation and Technology at the University of
Michigan--Flint, where he also serves as Accreditation Director and as
Special Assistant to the Vice Provost for Academic Strategy and
Effectiveness. He is an ABET Computing Accreditation Commission (CAC)
Commissioner and Team Chair. His research interests include
cybersecurity, AI/ML applied to security problems, wireless and
smart-grid security, and trust management, with a focus on the
intersection of security and emerging communication systems.
\end{IEEEbiographynophoto}

\end{document}